\documentclass[]{aa}
\usepackage{graphicx}
\usepackage{adjustbox}
\usepackage{txfonts}
\usepackage{booktabs} 
\usepackage{caption}
\usepackage{xcolor}
\usepackage{hhline}
\usepackage{multirow}
\usepackage{geometry}
\usepackage{tikz}

\makeatletter
\renewcommand*\aa@pageof{, page \thepage{} of \pageref*{LastPage}}
\makeatother
\usepackage{hyperref}
\hypersetup{
    colorlinks=true,
    linkcolor=blue,
    filecolor=magenta,      
    urlcolor=blue,
    citecolor=blue,
    }

\begin{document} 

\title{Tracing $\omega$Centauri's origins: Spatial and chemical signatures of its formation history}
\subtitle{}

\author{E. Dondoglio\inst{1},
A. P. Milone\inst{1,2},
A. F. Marino\inst{1},
A. Mastrobuono-Battisti\inst{1},
E. Bortolan\inst{2},
M. V. Legnardi\inst{2},
T. Ziliotto\inst{2},
F. Muratore\inst{2},
G. Cordoni\inst{3},
E. P. Lagioia\inst{4},
M. Tailo\inst{1}
}

\institute{
Istituto Nazionale di Astrofisica - Osservatorio Astronomico di Padova, Vicolo dell’Osservatorio 5, Padova, IT-35122 \\ \email{emanuele.dondoglio@inaf.it}
\and
Dipartimento di Fisica e Astronomia ``Galileo Galilei'', Univ. di Padova, Vicolo dell'Osservatorio 3, Padova, IT-35122
\and
Research School of Astronomy and Astrophysics, Australian National University, Canberra, ACT 2611, Australia
\and
South-Western Institute for Astronomy Research, Yunnan University, Kunming 650500, People's Republic of China
}

\titlerunning{Tracing $\omega$Centauri's Origins}
\authorrunning{Dondoglio et al.}

\date{Received XXX / Accepted XXX}

\abstract{

$\omega$Centauri ($\omega$Cen) is the most enigmatic Galactic globular cluster (GC), with unmatched chemical complexity. We combine photometric and spectroscopic catalogs to identify distinct stellar populations within $\omega$Cen, to investigate their spatial distribution and chemical properties, uncovering new insights into the cluster’s formation history.
Our population tagging identify the iron-poor stars commonly found in most GCs: the first population (1P), with halo-like chemical composition, and the second population (2P), enriched in elements produced by proton-capture processes. Similarly, we divided the iron-rich stars (the anomalous stars) into two groups: the AI and the AII, which exhibit light-element abundance distributions similar to 1P and 2P stars, respectively.
The wide radial extension of our dataset (five times the half-light radius), allowed to directly and unambiguously compare, for the first time, the fraction of these populations at different radii.
We find that 2P and AII stars are more centrally concentrated than the 1P and AI. The remarkable similarities between the 1P-2P and AI-AII radial distributions strongly suggests that these two groups of stars originated from similar mechanisms.
Our chemical analysis indicates that the 1P and AI stars (the lower stream) developed their inhomogeneities through core-collapse supernova (and possibly other massive stars' ejecta) self-enrichment, as supported by their increasing $\alpha$-element abundances with [Fe/H]. These populations contributed p-capture-processed material to the intracluster medium, from which the chemically extreme 2P and AII stars (the upper stream) formed. Additional polluters—such as intermediate-mass asymptotic giant branch stars and Type Ia supernovae—likely played a role in shaping the AII population.
Finally, we propose that 2P and AII stars with intermediate light-element abundances (the middle stream) formed via dilution between the pure ejecta that created the upper stream and lower-stream material.
}

\keywords{Techniques: photometric - Stars: abundances - Stars: Population II - Globular Clusters: general.}

\maketitle

\section{Introduction}
\label{sec:intro}

Decades of research have revealed that most massive, old Galactic Globular clusters (GCs) exhibit star-to-star chemical inhomogeneities, particularly in light elements, leading to well-known C–N, O–Na, and Mg–Al abundance anticorrelations \citep[e.g.,][]{marino2008, carretta2009, meszaros2020}. These patterns arise because, in addition to stars with halo-like chemical compositions (first population; 1P), GCs host second-population (2P) stars enriched in He, N, Na, and Al, and depleted in C, O, and Mg \citep[see reviews by][]{bastian2018, gratton2019, milone2022}.

The picture is complicated by the presence of further stellar populations in one-fifth of the studied Galactic GCs (typically referred as 'anomalous', opposite to the 'canonical' 1P and 2P stars) enriched in total C+N+O, [Fe/H], and/or s-process elements, the so-called Type II GCs in opposite of the Type I that lack these additional stars \citep[e.g.,][]{marino2009, yong2015, milone2017, dondoglio2023}.
While different scenarios have been proposed to explain the chemical profiles of these objects -such as prolonged star formation \citep[e.g.,][]{dantona2016} and merging of different GCs \citep[e.g.,][]{bekki2016, mastrobuono2019, calamida2020}- their origin is yet to be unveiled.

Among the already exceptional Type II GCs, $\omega$Centauri ($\omega$Cen) stands out as the most complex and peculiar case. Numerous studies have revealed its extraordinary chemical diversity, including a broad [Fe/H] spread exceeding 1 dex with distinct peaks in its distribution \citep[e.g.,][]{norris1995, johnson2010, nitschai2024}, and multiple, well-separated C–N, O–Na, and Mg–Al anticorrelations at different metallicities \citep[e.g.,][]{marino2011, meszaros2021, alvarez2022}.
The amplitude of light-element variations in $\omega$Cen is among the largest seen in any GC, likely linked to its large mass, as such variations correlate with cluster mass \citep{milone2018, meszaros2020, dondoglio2025}. Beyond iron and light elements, $\omega$Cen also shows strong s-process and total C$+$N$+$O enhancement, both of which increase with [Fe/H] and exceed the levels observed in any other Type II GC \citep[][]{norris1995, smith2000, johnson2010, marino2011, marino2012, meszaros2021, mason2025}.

The remarkable chemical complexity of $\omega$Cen indicates the presence of numerous distinct populations. Photometric studies revealed more and more sequences of stars as the quality of the observation increased. Early works by \citet{lee1999} and \citet{bedin2004} identified split sequences in color–magnitude diagrams (CMDs) along the main sequence (MS), sub-giant branch (SGB), and red-giant branch (RGB). \citet{bellini2017} used UV and optical {\it{HST}} data to detect at least 15 distinct populations in the MS. A major breakthrough came with the introduction of the chromosome map (ChM)—a pseudo two-color diagram constructed from the F275W, F336W, F438W, and F814W {\it{HST}} filters \citep{milone2015}—which enabled unprecedented populations tagging. Using the ChM, \citet{milone2017} identified 1P, 2P, and anomalous RGB stars in $\omega$~Cen's central 2 arcmin. More recently, \citet{haberle2024} extended this capability to $\sim$5 arcmin, expanding the radial coverage for ChM-based population tagging.

Radially extended populations tagging are instrumental to derive their spatial distribution —an essential aspect in GC studies, as the observed population gradients \citep[e.g.,][]{milone2012a, lee2017, leitinger2023} are closely tied to both formation mechanisms and dynamical evolution \citep[][]{mastrobuono2016, dalessandro2019, vesperini2021}.
Previous works traced the radial distribution of multiple distinct MSs \citep{sollima2007, bellini2009, scalco2024} out to $\sim$20 arcmin from the center, revealing that these populations are not fully mixed, with some more centrally concentrated than others. However, in a system as complex as $\omega$Cen, CMD-based MS separation does not allow an unambiguous assignment of these sequences to 1P, 2P, or anomalous populations, limiting a comparison with the spatial distributions in other GCs.

The origin of $\omega$Cen’s complex structure has long been debated. A widely accepted scenario suggests that it is the remnant nucleus of a more massive stellar system—likely a dwarf galaxy—whose outer regions were stripped during accretion by the Milky Way, as first modeled by \citet{bekki2003}. Supporting this view, \citet{pagnini2025} pointed out that other GCs show chemical similarities with $\omega$Cen, implying a common origin within the same progenitor galaxy. The idea of $\omega$Cen being a nuclear star cluster of this progenitor galaxy has been often considered, with this GC resulting from a mix of in-situ star formation and infall of other GCs \citep{neumayer2020}.
On the other hand, other works suggested that the chemical complexity derived from self enrichment due to a prolonged star formation, without invoking any merger events \citep[e.g.,][]{pancino2007, johnson2010}.

Despite this, the detailed origin of its intricate light- and heavy-element abundance patterns remains uncertain. \citet{alvarez2024} identified a discontinuity in the Mg–Al trend, suggesting multiple formation channels for light-element inhomogeneities at different metallicities. 
\citet{mason2025} proposed that different groups of stars evolved separately in GC-like environment and then merged to assemble $\omega$Cen. Conversely, \citet{marino2012} presented a scenario rooted in the self-enrichment framework previously proposed for all GCs \citep[e.g.,][]{prantzos2006}, but extended to account for the complexity of this particular system. Here, 1P-like stars formed first, followed by rapid starbursts that created heavy-element inhomogeneities, leading to the formation of subsequent generations with light-element variations via self-pollution.
Another ongoing debate, tightly linked with its formation history, concerns $\omega$Cen’s age spread. Several studies, based on CMD analysis of the SGB, have suggested an age spread of up to $\sim$2 Gyr \citep[e.g.,][]{lee1999, villanova2014, clontz2024}. However, \citet{tailo2016}, using multi-band {\it{HST}} data and stellar models, concluded that all stars likely formed within a much shorter period, not exceeding 0.5 Gyr.
A key source of uncertainty in age determinations is the C$+$N$+$O variation, which can mimic age differences by producing similar SGB splits. Accurately constraining the total C$+$N$+$O content is thus essential to resolve the age spread controversy.

In this work, we consistently identify 1P, 2P, and anomalous stars in $\omega$Cen from its center out to $\sim$30 arcmin (approximately five times the half-light radius) for the first time, by combining multiple datasets—both photometric and spectroscopic—to investigate their spatial distribution and chemical composition.
Section~\ref{sec:2} describes the datasets used in this study. In Section~\ref{sec:3}, we apply these data to identify the same populations across different datasets over a wide radial range. Section~\ref{sec:4} examines the radial distribution of the identified populations.
Section~\ref{sec:5} explores the chemical composition of the different stellar components, while Section~\ref{sec:6} presents our reconstruction of the formation sequence that shaped the cluster. Finally, our main results and conclusions are summarized in Section~\ref{sec:7}.

\section{Dataset}
\label{sec:2}

We combine photometric and spectroscopic datasets of $\omega$Cen to identify distinct stellar populations along the RGB across a wide range of radial distances from the cluster center.

We utilize three photometric catalogs in our analysis. First, we employ the dataset and the ChM published by \citet{milone2017}, constructed using UV and optical {\it{HST}} photometry in the F275W, F336W, F438W, and F814W filters, covering the innermost $\sim$2 arcmin of the cluster.
Second, we use the {\it{HST}} catalog by \citet{haberle2024}, which includes multi-epoch {\it{HST}} and optical photometry in bands suitable for constructing a ChM. This catalog combines decades of observations across different fields of view, extending up to $\sim$5 arcmin from the center. The photometry was corrected for differential reddening following the procedure described by \citet{milone2012b}.
Third, we incorporated Gaia photometry \citep{gaia2023} to assess membership for stars located beyond the 5 arcmin radius not included in the previous two {\it{HST}}-based catalogs. We selected $\omega$Cen members based on high-quality Gaia photometry\footnote{We used the renormalized unit weight error \citep[{\tt{RUWE}}; e.g.,][]{lindegren2018} as high photometric quality diagnostic.}, proper motions and parallaxes consistent with GC stars, following the approach of \citet{cordoni2018} and \citet{jang2022}.

\begin{figure*}
\includegraphics[width=8.5cm, clip, trim={ 0cm 0cm 17.7cm 0cm}]{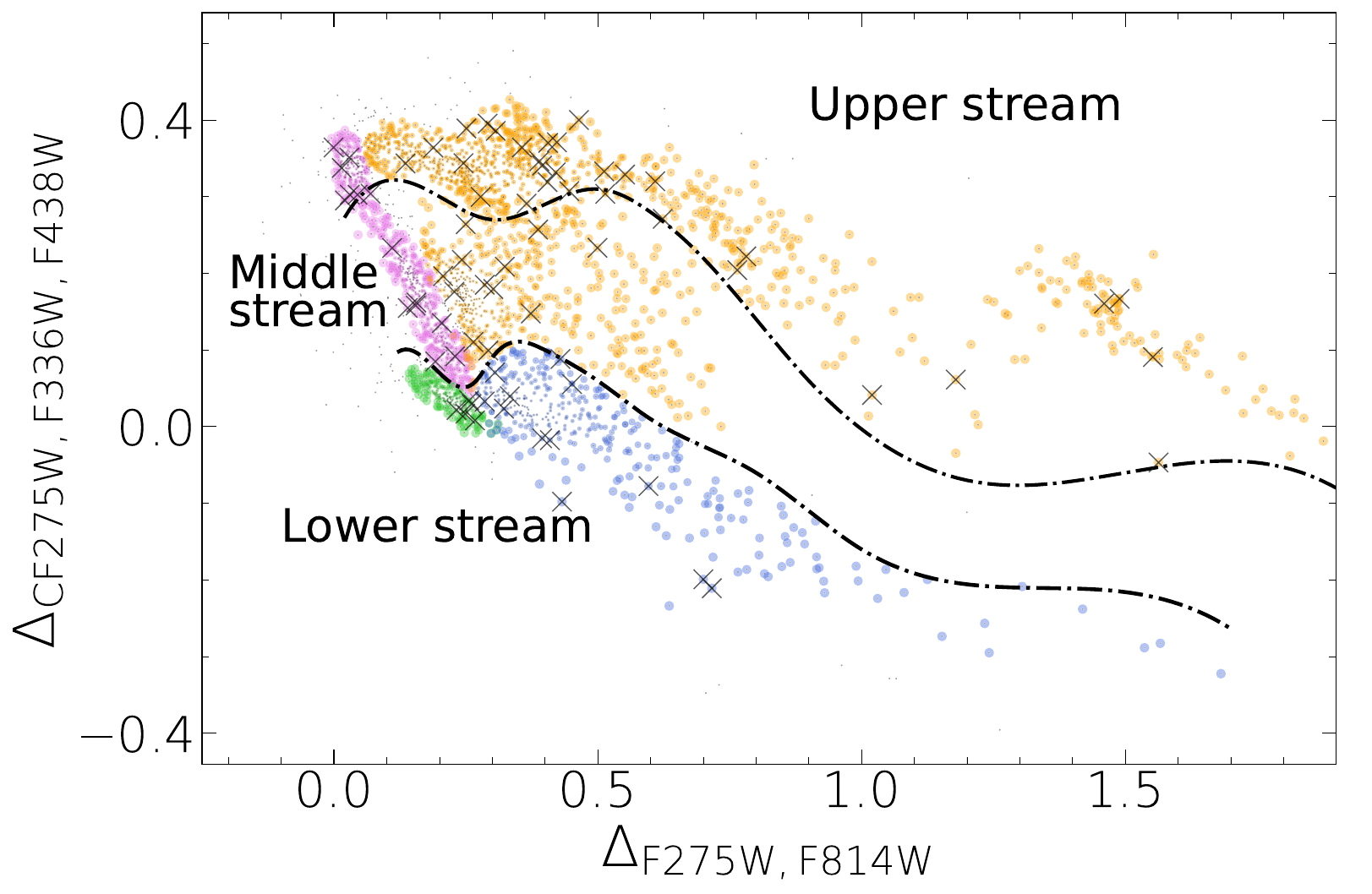}
\includegraphics[width=8.5cm, clip, trim={ 0cm 0cm 17.7cm 0cm}]{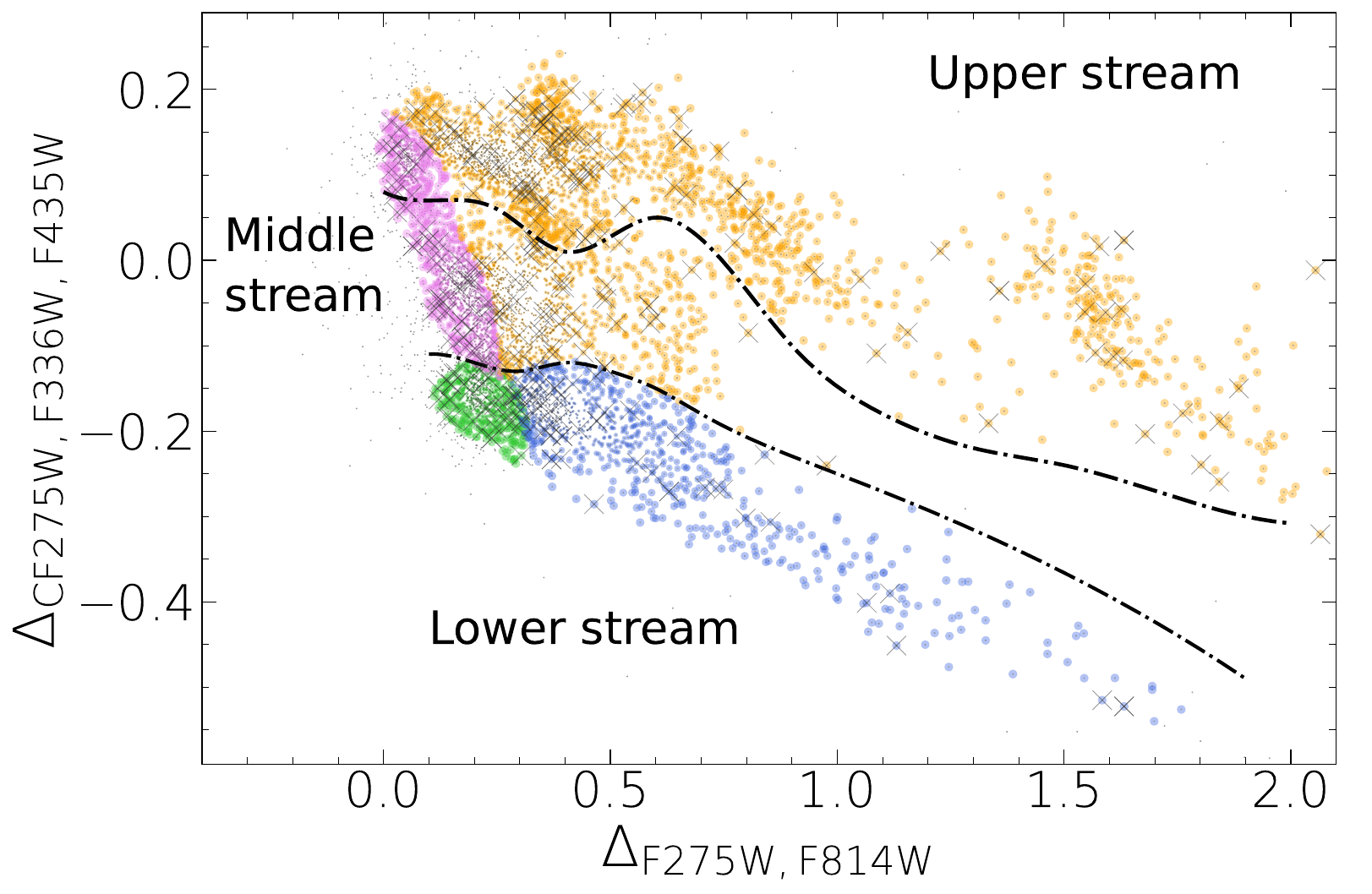}
\caption{ChM of the innermost two arcmin of $\omega$Cen published by \citep{milone2017} (left panel) and made from the {\it{HST}} catalog from \citet{haberle2024} covering the 2-5 arcmin range (right panel). 1P, 2P, AI, and AII stars are colored in green, violet, blue, and orange, respectively. The two dot-dashed black lines separates the lower, middle, and upper streams as reported in each plot.
The black crosses indicate stars that are present in our spectrosocpic dataset presented in Section~\ref{sec:2}.}
\label{fig:chms_hst}
\end{figure*}

Our primary spectroscopic dataset is drawn from APOGEE Data Release 17 \citep[DR17;][]{abdurro2022}, from which we selected only bright stars with signal-to-noise ratios $>$70, and excluded sources flagged with {\tt{ASCAPFLAG}} = {\tt{STAR\_BAD}}\footnote{APOGEE documentation at {\url{https://www.sdss4.org/dr17/}}.}. For a comprehensive overview of our selection criteria, we refer the reader to \citet{dondoglio2025}. In this study, we consider the following abundance ratios from APOGEE: [C/Fe], [N/Fe], [O/Fe], [Mg/Fe], [Al/Fe], [Si/Fe], [Ca/Fe], [Fe/H], and [Ce/Fe].
Additionally, we include two other datasets. From \citet{marino2011}, we use measurements of [Na/Fe], [Ba/Fe], and [La/Fe], and from \citet{mucciarelli2018}, we employ A(Li) abundances.
Across all spectroscopic datasets, we retained only stars whose according to photometric catalogs and radial velocities are consistent with $\omega$Cen membership. This filtering also allows us to exclude [C/Fe] and [N/Fe] measurements in stars brighter than the RGB bump, which are affected by extra mixing processes after the first dredge-up, leading to carbon depletion and nitrogen enhancement depending on luminosity \citep[e.g.,][]{charbonnel1998, shetrone2019, lee2023}, potentially biasing our interpretation of chemical differences among cluster populations. Although lithium is also affected by similar processes, we did not apply this additional selection to A(Li), as the dataset from \citet{mucciarelli2018} already includes stars located below the RGB bump only.
The spectroscopic dataset allows us to get abundances of $\omega$Cen stars up to about 30 arcmin from the cluster center.

\section{Population Tagging}
\label{sec:3}

We dedicate this Section to identify the different populations that will be analyzed in this work by exploiting multiple photometric and spectroscopic diagrams to spot them from the innermost area up to about 30 arcmin from the cluster center.
Section~\ref{sec:3.1} investigates the ChMs built with {\it{HST}} photometry, while Section~\ref{sec:3.2} is dedicated to tag these populations through APOGEE spectroscopic data.

\subsection{Populations in the Chromosome Map} \label{sec:3.1}

The complexity of the morphology of $\omega$Cen is unmatched by any other GC. This is clearly represented by the $\Delta_{\rm CF275W,F336W,F438W}$ vs. $\Delta_{\rm F275W,F814W}$ ChM of RGB stars in the innermost $\sim$2 arcmin published by \citet{milone2017}, represented in the left panel of Figure~\ref{fig:chms_hst}. Here, multiple-populations describes several distinct blobs of stars, as well as gaps and extended sequence.
For that, classifying $\omega$Cen population is far from an easy task, and several classification models have been proposed throughout the years. 
Milone and collaborators divide the stars in the ChM between the canonical, that form the 1P and 2P pattern typical of all GCs, and the anomalous, which are characterized by iron, s-process, and C+N+O enhancement, only present in Type II GCs.
Following the prescription by \citet{milone2017}, canonical stars form the bluest sequence on this diagram, with 1P stars forming a distinct clump at around $\Delta_{\rm F275W,F336W,F438W} \sim$0 in the ChM, colored in green in Figure~\ref{fig:chms_hst}. The remaining sequence of stars, which span a $\Delta_{\rm CF275W,F336W,F438W}$ range between $\sim$0.1-to-0.4, is made by 2P stars (in violet).

Now, we divide the anomalous stars into two subgroups by following a similar approach: one population that describes a sequence with low $\Delta_{\rm CF275W,F336W,F438W}$ (AI, in blue), and the second that includes all the other anomalous stars (AII, in orange). The position on the ChM suggests that AI does not exhibit nitrogen enhancement (primarily traced by the ChM y-axis) indicating that these stars did not form from material enriched by p-capture processes. On the contrary, the ChM position of AII stars suggests that they display signs of 2P-like chemical composition, as also suggested by spectroscopy \citep[e.g.,][]{johnson2010, marino2011, alvarez2024}.

In the right panel of Figure~\ref{fig:chms_hst}, we show the same population tagging on the $\Delta_{\rm CF275W,F336W,F435W}$ vs. $\Delta_{\rm F275W,F814W}$ ChM, obtained by applying the procedure of \citet[][see their Appendix A]{milone2017} to RGB stars in the $\omega$Cen catalog recently published by \citet{haberle2024}. This catalog reaches higher distances from the cluster center, allowing us to trace 1P, 2P, AI, and AII populations out to 5 arcmin. In both ChMs, we indicate with black crosses the stars that are part of the spectroscopic dataset introduced in Section~\ref{sec:2}.

Before moving on, we mention another popular classification that we will exploit in this work, which divide $\omega$Cen stars into three groups based on their position along the ChM y-axis: the lower, upper, and middle stream \citep[][]{marino2019, clontz2024}. The lower stream includes all the stars with no signs of p-capture pollution, so it coincides with 1P+AI, while the upper stream is made by the most chemically extreme 2P and AII stars, characterized by the largest light-element inhomogeneities (and the largest ChM y-axis coordinates). Finally, the middle stream consists in all the 2P and AII stars with intermediate ChM position. The black dash-dotted lines in both panels of Figure~\ref{fig:chms_hst} divide the three streams following the classification introduced by \citet{marino2019}. Table~\ref{tab:classify} provides a summary on the different classification criteria and how they relate to each others.

\subsection{APOGEE data in the outer regions} \label{sec:3.2}

Beyond five arcmin, no ChMs are available to identify 1P, 2P, AI, and AII RGB stars. To overcome this limitation, we used an alternative approach based on combining spectroscopic abundances provided by the APOGEE survey with Gaia photometry.

\begin{figure}
\includegraphics[width=6.8cm, clip, trim={0cm 0cm 0cm 0cm}]{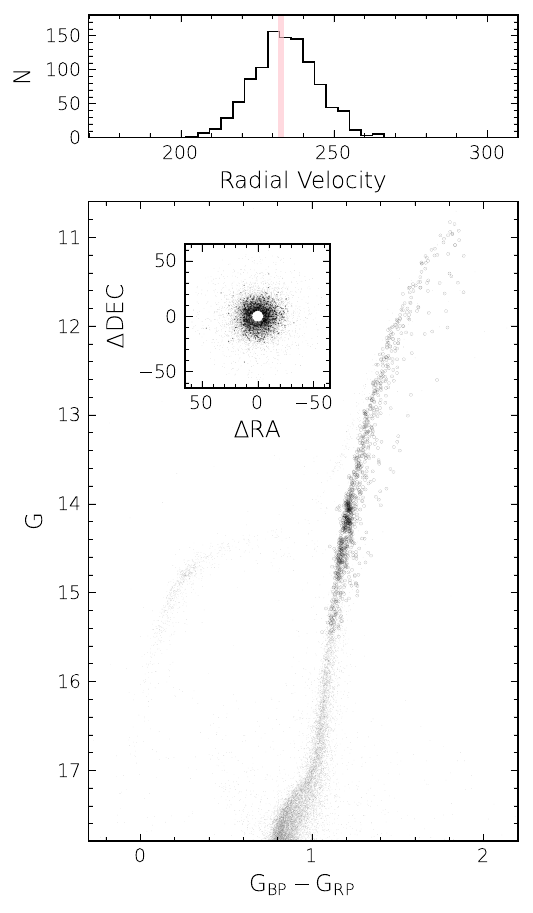}
\caption{$G$ vs. $G_{\rm BP}$-$G_{\rm RP}$ from Gaia photometry. Grey points represent $\omega$Cen members with high-quality photometry, while black dots indicate the RGB stars with available APOGEE abundances, who reach up to $\sim$30 arcmin radius. The inset shows the $\Delta$DEC vs. $\Delta$RA position respect to the cluster center (in arcmin) of the stars shown in the CMD. The top panel displays the radial velocity histogram of the selected sample of stars.}
\label{fig:gaia}
\end{figure}

The first step consist in assessing a sample of $\omega$Cen RGB stars outside a radius of five arcmin. To do that, we exploit the Gaia $G$ vs. $G_{\rm BP}$-$G_{\rm RP}$ CMD of $\omega$Cen members (selected as in Section~\ref{sec:2}) illustrated in Figure~\ref{fig:gaia} in a distance range between five and 60 arcmin from the cluster center as shown in the inset panel. The stars highlighted with black dots represent RGB stars with available APOGEE abundances\footnote{We excluded AGB stars based on their position in the CMD.}. In the top panel, we show the histogram of the radial velocity values for our sample of stars, which is distributed around the average radial velocity (pink line) reported by \citet{baumgardt2018}.

\begin{figure}
\includegraphics[width=7.5cm, clip, trim={0cm 0cm 0cm 0cm}]{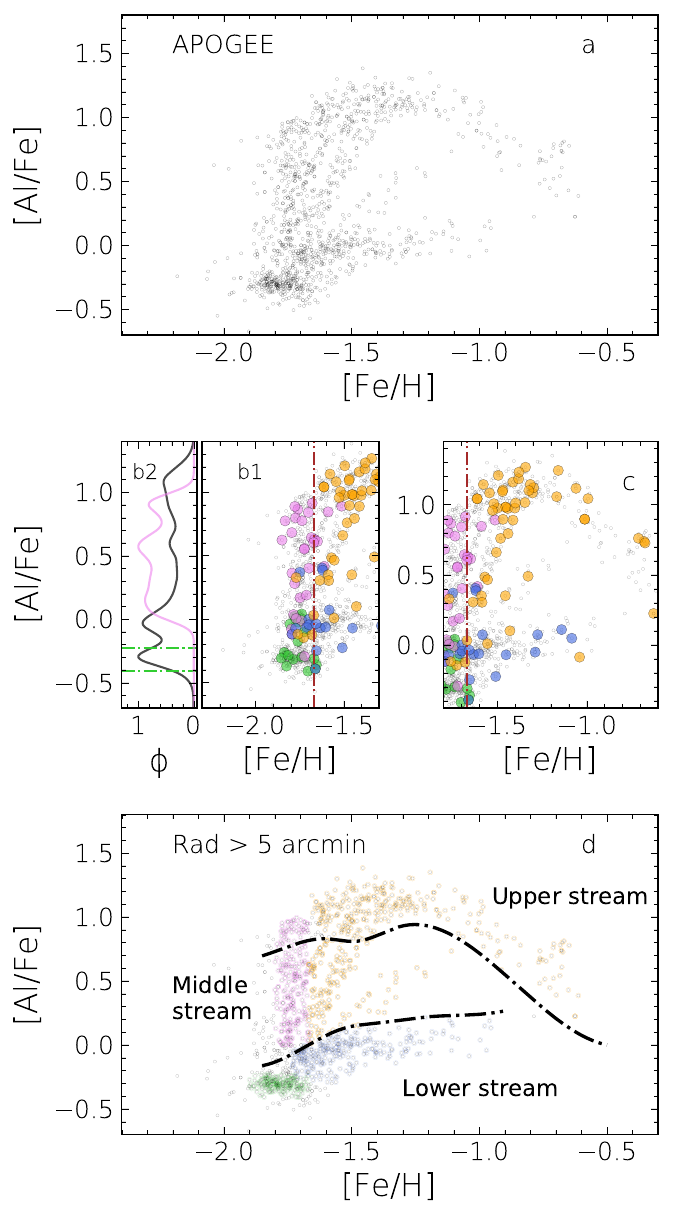}
\caption{{\it{Panel a:}} APOGEE [Al/Fe] vs. [Fe/H] for RGB stars of $\omega$Cen.
{\it{Panel b1:}} Zoom-in around the iron-poor end. Stars with ChM tagging are colored according to their assigned population, while the vertical dot-dashed line indicates the 90th percentile of the 1P stars [Fe/H] distribution.
{\it{Panel b2:}} normalized kernel density distribution of [Al/Fe] of all the stars (black) and of 2P stars (violet). The two green dot-dashed horizontal lines indicate the 1P stars [Al/Fe] range. {\it{Panel c:}} same as panel b1 but zoomed at higher [Fe/H]. {\it{Panel d:}} [Al/Fe] vs. [Fe/H] of APOGEE stars outside 5 arcmin, colored according to the result of the population tagging performed in Section~\ref{sec:3.2}. The black lines separate the three streams, as reported in the plot. }
\label{fig:al_fe_pop}
\end{figure}

Our second step consists in finding the counterparts of the 1P, 2P, AI, and AII populations. To do that, we exploit the [Al/Fe] vs. [Fe/H] diagram of APOGEE abundances of all the RGB stars from the center to 30 arcmin, presented in panel a of Figure~\ref{fig:al_fe_pop}. This plot resembles in several ways the ChM, showing a continuous distribution of stars at [Fe/H]$<$-1.6 dex, populated by the canonical stars, and two separated elongated sequences at larger iron values, which are part of the upper and lower streams. The middle stream (with intermediate [Al/Fe]) is almost absent at larger [Fe/H], which is akin to what can be observed in the ChM at larger $\Delta_{\rm F275W,F814W}$. 
Panel b1 displays a zoom-in in the most metal-poor area, where we expect to be dominated by the canonical populations. Here, we colored all the stars in the APOGEE catalog within five arcmin with available ChM population tagging. Unsurprisingly, the distinct blob at low [Al/Fe] ($\sim$-0.3 dex) is mostly populated by 1P stars, and form a separated peak in the [Al/Fe] kernel density distribution portrayed in panel b2 (black line). We consider as 1P stars the ones with [Al/Fe] within the two horizontal dot-dashed green lines, and with [Fe/H] smaller than the brown vertical line in panel b1, which indicates the 90th percentile of the [Fe/H] distribution of the stars within the aluminum peak.
Guided by the stars with known classification, we consider as 2P the ones with [Al/Fe] between -0.05 and 1.10 dex and [FeH] below the brown-line level (as 2P stars are not iron enriched compared to 1P stars). We display in panel b2 the [Al/Fe] kernel density distribution of 2P stars (violet line), in which a distinct peak above [Al/Fe]$\sim$0.8 dex is visible. We interpret this peak as the canonical part of the upper stream, i.e. the counterpart of the high-$\Delta_{\rm CF275W,F336W,F438W}$ 2P stars, which form a separate blob in the ChM.
Panel c illustrates a zoom-in focused on the anomalous stars, showing that the elongated sequence below [Al/Fe]$\sim$0.1 dex is populated by AI stars, while AII stars dominate the area above that level on the red side of the brown line, as expected from the distribution of these stars in the ChM.

Panel d presents the final result of our tagging, where stars in the [Al/Fe] vs. [Fe/H] diagram outside five arcmin are colored according to the assigned population. The two black dash-dotted lines separates, as in Figure~\ref{fig:chms_hst}, the lower, middle and upper streams. While the definition of the lower stream is straightforward, as is the combination of 1P and AI stars, separating middle and upper streams is not an obvious task along all the diagram. We separate these two groups by eye based on the fact that the 2P stars above $\sim$0.8 dex, as discussed in the previous paragraph, are very likely to belong to the upper stream, and that the middle stream tends to disappear at larger [Fe/H]. These facts, combined with the knowledge from spectroscopy that we expect the largest aluminum enhancement among upper stream stars at each [Fe/H] \citep[e.g.,][]{marino2019, mason2025}, ensures that our separation is overall reliable even in the [Al/Fe] vs. [Fe/H] plane.

\subsubsection{Binary stars} \label{sec:3.3.1}

The information present in the APOGEE dataset also allows us to identify candidate binary stars in our sample. To do that, we followed the approach outlined in the APOGEE documentation\footnote{{\tt{https://www.sdss4.org/dr17/irspec/radialvelocities/}.}} based on the {\tt{VSCATTER}} quantity, which indicates the scatter of radial velocity measured for each star in case of multiple measurements.
We identify as likely binaries the stars with {\tt{VSCATTER}} > 1 $km s^{\rm -1}$, thus with a scatter much larger than observational errors, as recommended in the documentation.

\begin{figure}
\includegraphics[width=7.5cm, clip, trim={11cm 0cm 0cm 0cm}]{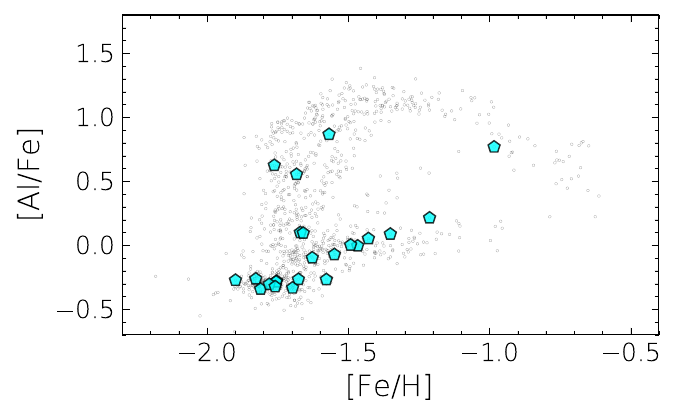}
\caption{[Al/Fe] vs. [Fe/H] diagram, as in Figure~\ref{fig:al_fe_pop}, where probable binaries are highlighted with cyan diamonds.}
\label{fig:binaries}
\end{figure}

We find 23 binaries that also had the population tagging performed in Figure~\ref{fig:al_fe_pop}, all with three to thirteen radial velocity measurement. We show them in Figure~\ref{fig:binaries} as cyan diamonds in the [Al/Fe] vs. [Fe/H] diagram. We identify 9 1P, 3 2P, 8 AI, and 3 AII binaries. The majority of stars (about 70\%) belong to the 1P and AI population in this radial range ($\sim$5.9-21.6 arcmin). Notably, if we consider the outermost half of stars (outside 14 arcmin), the incidence of 1P+AI binaries increases to about 90\%.

\begin{table}
    \centering
\caption{Different classification of $\omega$Cen populations (see text for details).}
\begin{tabular}{c|c|c|c}
    \hline
    \hline
     & & & \\
                  & Canonical & Anomalous & Stream \\
    & & & \\
    \hline
    & & & \\
    no p-capture  & 1P        & AI        & lower \\
    & & & \\
    yes p-capture & 2P        & AII       & middle + upper \\
    & & & \\
\hline
\hline
%    & & & \\
\end{tabular}
\label{tab:classify}
\end{table}

\subsubsection{Extended 1P sequence} \label{sec:3.3.2}

We end this Section by noticing that, from a visual inspection of the [Al/Fe] vs [Fe/H] diagram in Figure~\ref{fig:al_fe_pop}, the 1P stars appear distributed on a more extended [Fe/H] range than the 2P stars.
 This is not unexpected, as small iron spread among 1P stars have been observed in an increasing amount of GCs in the past years \citep[e.g.][]{marino2019a, legnardi2022, marino2023, dondoglio2025}. We infer the width of the iron spread among 1P stars by following the procedure exploited by \citet{dondoglio2025}. Briefly, we calculated the 90th$-$10th percentile of the 1P [Fe/H] values and subtracted to it in quadrature the same quantity measured from a simulated Gaussian distribution with sigma equal to the typical [Fe/H] observational error provided by the APOGEE dataset.

We found that the 1P width values 0.138$\pm$0.008, which is significantly larger than the 2P iron width (0.085$\pm$0.003). Moreover, we point out how our measurement from direct spectrosocpic information is in a 1-$\sigma$ level agreement with the amount of spread predicted by \citet{legnardi2022} from the ChM distribution of 1P stars (0.148$\pm$0.011).

\section{Spatial distribution}
\label{sec:4}

The population tagging described in Section~\ref{sec:3} enables an unprecedented analysis of the spatial distribution of RGB stars belonging to different populations within $\omega$ Cen across a wide radial range.

\begin{figure}
\includegraphics[width=7.5cm, clip, trim={0cm 0cm 0cm 0cm}]{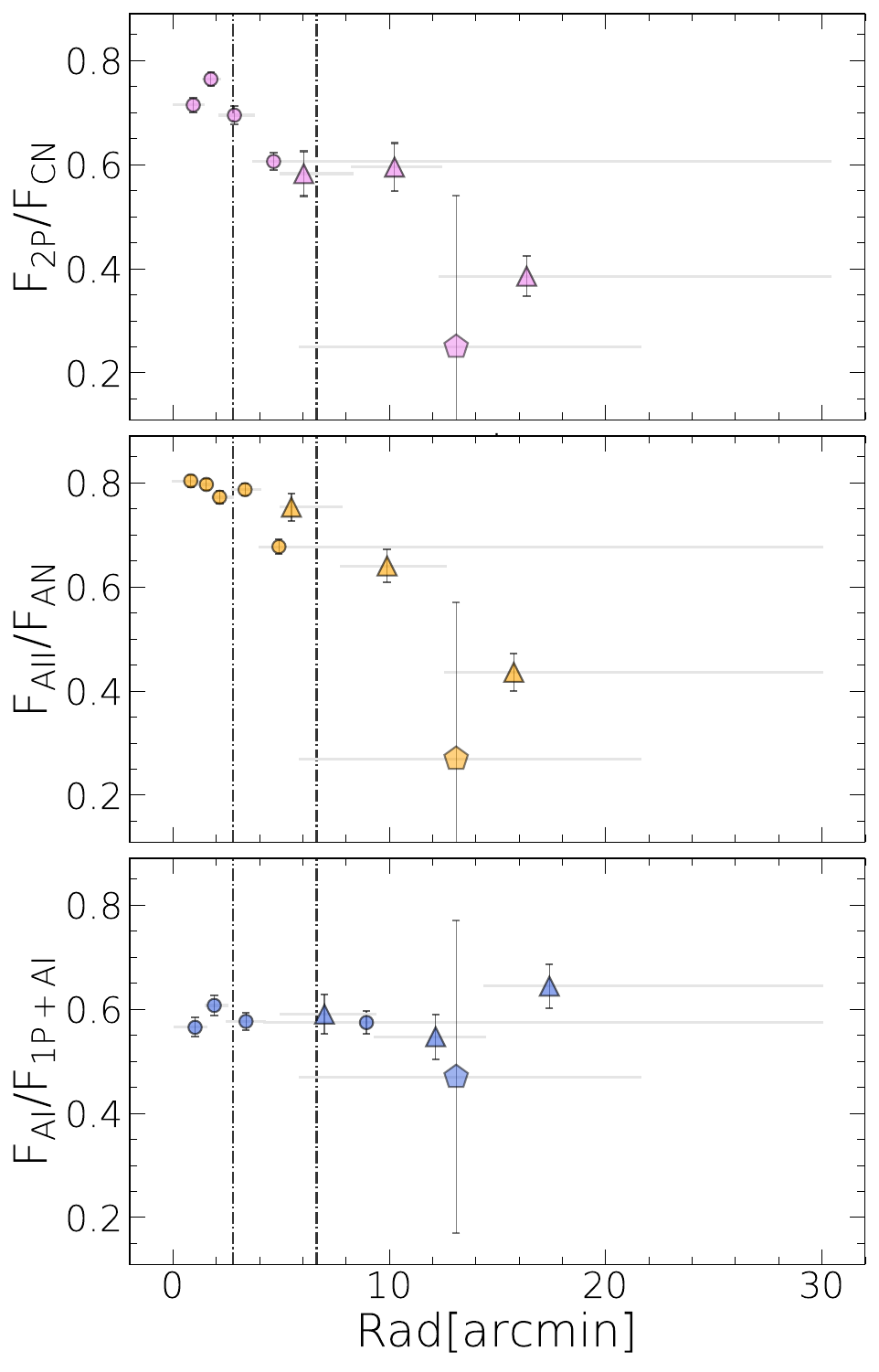}
\caption{Fraction of 2P respect to the total amount of canonical stars (top), AII compared to the bulk of anomalous (middle), and AI to the whole lower stream (bottom) against the radial distance from $\omega$Cen center. Bullets indicates fractions obtained from the whole dataset combined, while triangles are based on the APOGEE dataset only (see text for details). Diamonds represent the fraction from binaries stars only. The gray lines represent the radial interval relative to each measurement, while the dot-dashed black lines indicate the core and half-mass radius.}
\label{fig:radial}
\end{figure}

In the upper panel of Figure~\ref{fig:radial}, we present the radial distribution of the 2P stars fraction respect to the total number of canonical stars. Radial bins with equal numbers of stars are considered, first for the entire dataset (dots) and then for the outermost dataset -presented in Section~\ref{sec:3.2}- only (triangles), to increase the resolution in the less dense outer regions.
The two vertical dot-dashed lines represent the core and half-mass radius from \citet{baumgardt2018}.
The fraction of 2P stars decreases with distance from the center, nearly halving between the innermost and outermost regions (comparing values at the center and at 20 arcmin), indicating that 2P stars are more centrally concentrated than 1P stars.
The diamond indicates the fraction of 2P binary stars, which is smaller than the one observed for the non-binary stars in the same radial interval, reaching around 0.25.

The central panel displays similar calculations but for the AII population compared to the bulk of anomalous. The AII stars are more centrally concentrated than the AI, mirroring the pattern seen between 2P and 1P stars. Moreover, the radial distributions of 2P and AII (and, consequently, of 1P and AI) stars span similar ranges, with values shifting from approximately 0.75 in the innermost 3 arcmin to 0.40 beyond 10 arcmin. A further similarity consist in the AII binary fraction, which is smaller than the AII non-binary fraction at similar radial range.

The lower panel shows the fraction of AI stars relative to the total number of 1P and AI (i.e., the bulk of the lower stream). This ratio remains relatively constant ($\sim$0.6) over the 30 arcmin range, with no evident differences with the binary fraction.

We tested the statistical significance of the observed radial trends using a Kolmogorov–Smirnov test to obtain a p-value, which measures the likelihood that the observed distributions are the result of random statistical fluctuations, rather than representing intrinsic variations. The distributions in the upper and middle panels of Figure~\ref{fig:radial} have p-values less than 0.01, supporting the hypothesis that the observed radial variations are physical. In contrast, the fraction of AI stars relative to 1P stars has a p-value of 0.60, strongly suggesting that the flat distribution is indeed intrinsic. Overall, the p-value test confirms the trends visually inspected in this section.

This is the most comprehensive and unequivocal identification of 1P and 2P stars over such a wide radial  range in $\omega$Cen, providing a robust confirmation of previous results of the 2P stars segregation \citep[e.g.,][]{sollima2007, bellini2009, johnson2010, scalco2024}.
The fact that the fraction of 1P binaries is higher compared to the bulk of stars favors this idea: since 2P stars formed in a more centrally concentrated environment, their disruption rate is higher than the more diffused 1P \citep[e.g.,][]{vesperini2011, hypki2022}, leading to smaller amount of 2P binaries, as also observed in other GCs \citep[e.g.,][]{dorazi2010, lucatello2015, dalessandro2018, bortolan2025}.
Interestingly, AI and AII stars exhibit both a radial distribution and a binary incidence similar to those of 1P and 2P stars, respectively. This strongly suggests that the 2P and AII populations formed through similar mechanisms, after the 1P and AI stars were already developed.

\section{Exploring the chemical composition of $\omega$Cen}
\label{sec:5}

Our finding in Section~\ref{sec:4} about the similarities between the 1P and AI, and 2P and AII spatial distribution suggests that these two pairs of population share a similar formation origin.
To further explore this idea, we dedicate this Section to investigate the chemical composition of the different populations identified in $\omega$Cen, with particular focus on the three streams. Section~\ref{sec:5.1} compares the chemical composition of the lower and upper streams, while Section~\ref{sec:5.2} focuses on the middle stream.

\begin{figure*}
\includegraphics[width=18cm, clip, trim={0cm 0cm 0cm 0cm}]{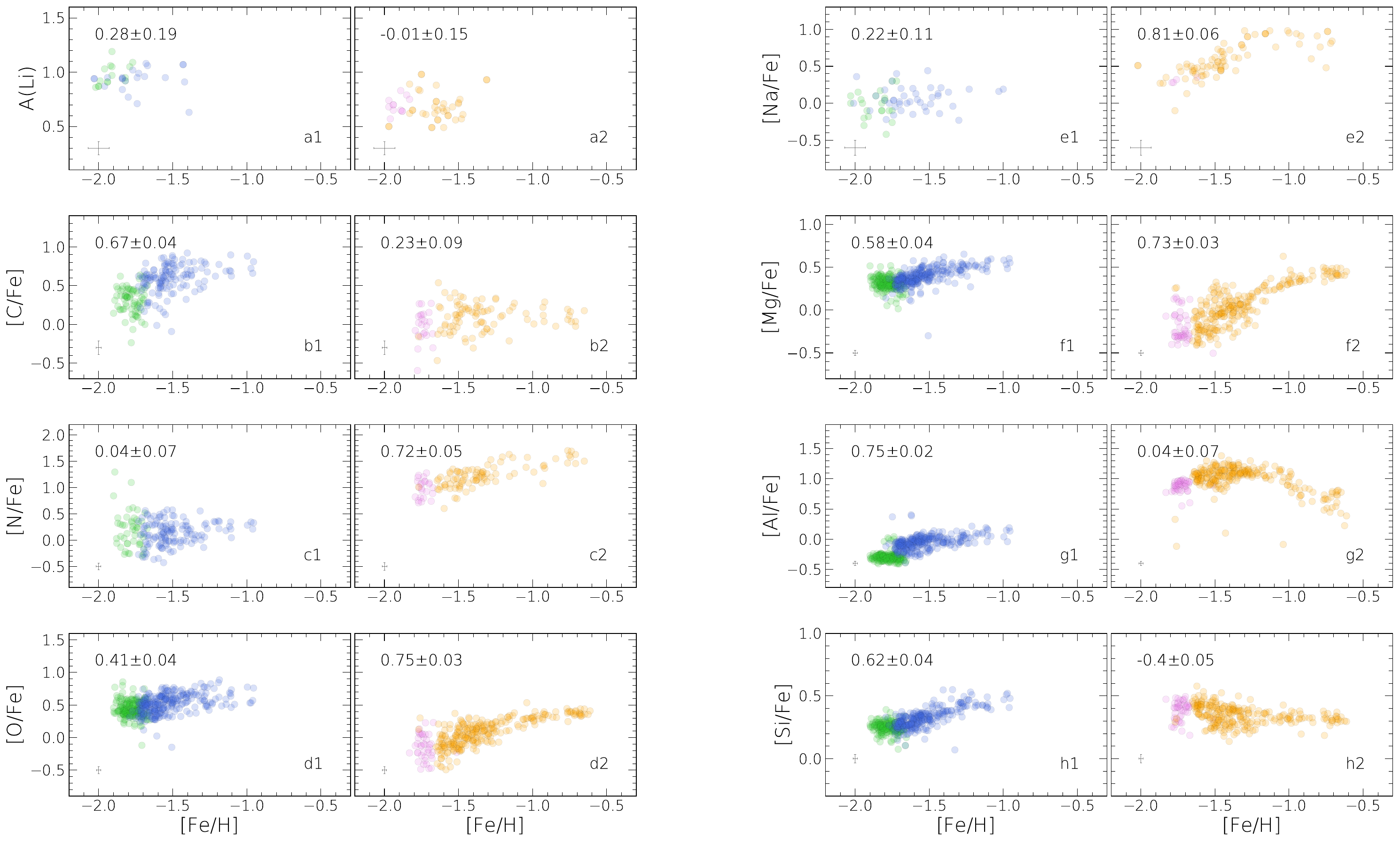}
\caption{A(Li), [C/Fe], [N/Fe], [O/Fe], [Na/Fe], [Mg/Fe], [Al/Fe], and [Si/Fe] vs [Fe/H] of the lower stream (panels a1-h1) and upper stream (panels a2-h2) stars. Green, purple, violet, and orange dots indicate 1P, AI, 2P, and AII stars, respectively. The numbers reported in the panels illustrate the Spearman's correlation coefficient and its associated error. }
\label{fig:chem_ls_us}
\end{figure*}

\subsection{Comparing the lower and upper streams}
\label{sec:5.1}

In Figure~\ref{fig:chem_ls_us}, we examine the distributions of light-element abundances as a function of [Fe/H] for stars in the lower and upper streams, shown in panels a1–h1 and a2–h2, respectively. The abundances considered —A(Li), [C/Fe], [N/Fe], [O/Fe], [Na/Fe], [Mg/Fe], [Al/Fe], and [Si/Fe]— are all known to differ between the two streams \citep[e.g.,][]{johnson2010, marino2011, mucciarelli2018, marino2019, alvarez2024, mason2025}. Stars are color-coded according to their association with the 1P, 2P, AI, and AII populations.
Upper stream stars, on average, exhibit smaller lithium, carbon, oxygen, and magnesium, and larger values of nitrogen, sodium, and aluminum. These differences are broadly consistent with the upper stream 
likely originating from a material polluted by the high mass stars' ejecta of the lower stream population.

However, intrinsic abundance variations with [Fe/H] complicate the interpretation of these trends. To explore potential correlations, we computed Spearman’s rank correlation coefficient ($R_{\rm s}$), reported in each panel. The associated uncertainties were derived from the standard deviation of 1,000 resamplings, with data points scattered according to their observational errors.
In the lower stream, [C/Fe], [O/Fe], [Mg/Fe], [Al/Fe], and [Si/Fe] show positive correlations with [Fe/H], whereas A(Li), [N/Fe], and [Na/Fe] appear roughly constant. The upper stream displays both similarities and key differences: A(Li), [O/Fe], and [Mg/Fe] follow trends comparable to those in the lower stream, but other elements diverge. In particular, [C/Fe] remains constant, while [Na/Fe] and [N/Fe] increase sharply with [Fe/H]. The [Al/Fe] trend in the upper stream describes an arc-like structure, peaking around [Fe/H]~$\sim$-1.6 dex before declining at higher metallicities. Notably, [Si/Fe] in the upper stream anticorrelates with [Fe/H], opposite to the trend observed in panel h1. Consequently, silicon is higher than in the lower stream only in the low-metallicity end, with this trend inverting towards higher [Fe/H].

The increase in [O/Fe], [Mg/Fe], and [Si/Fe] among lower stream stars points toward an overall enhancement of $\alpha$-elements with increasing [Fe/H]. To further investigate this, we calculated the mean [$\alpha$/Fe] for lower stream stars by averaging their [O/Fe], [Mg/Fe], [Si/Fe], and [Ca/Fe] abundances.
The [Ca/Fe] vs. [Fe/H] trend of the lower stream is represented in the upper panel of Figure~\ref{fig:chem_alpha}. For completeness, we also present the same plot for the upper stream in the middle panel. The lower panel displays the resulting [$\alpha$/Fe] for the lower stream, which, as expected, increases with iron content.

Finally, we turn our attention to the s-process elements and the total [C+N+O/Fe], which are known to vary significantly in $\omega$Cen \citep[e.g.,][]{marino2011, meszaros2021}. Figure~\ref{fig:chem_agb} presents the behavior of the lower and upper streams for the available s-process element measurements, including [Ce/Fe] from APOGEE (panels a1 and a2) and the average of [Ba/Fe] and [La/Fe] ($[{\rm \langle Ba,La\rangle}/\rm Fe]$, panels b1 and b2) from \citet{marino2011}.
[Ce/Fe] clearly increases with [Fe/H], as confirmed by the corresponding $R_{\rm s}$ values. A similar trend is evident in the barium and lanthanum average, albeit derived from a smaller sample. Interestingly, at low metallicity, the two streams exhibit similar s-process abundances, but at higher [Fe/H], the upper stream displays higher [Ce/Fe], as first reported by \citet{mason2025}.

The comparison between the lower and upper streams underscores the complexity of interpreting their origins. Although a scenario in which the lower stream formed first and enriched the intracluster medium—leading to the formation of the upper stream—is consistent with the observed chemical differences, it cannot fully explain the morphologies presented in Figures~\ref{fig:chem_ls_us},~\ref{fig:chem_alpha}, and~\ref{fig:chem_agb}. We will explore this issue in more depth in Section~\ref{sec:6}.

\subsection{The intermediate nature of the middle stream}
\label{sec:5.2}

The middle stream is composed of 2P and AII stars whose positions along the ChM y-axis and [Al/Fe] values fall between the upper and lower streams. Its most distinctive feature is a narrower [Fe/H] distribution compared to the other two streams, and is more commonly found among canonical stars but becomes increasingly rare at higher [Fe/H] (and ChM x-axis) values.

The intermediate chemical behavior of the middle stream is illustrated in Figure~\ref{fig:chem_middle}, where we show the median values for middle stream 2P and AII stars in violet and orange, respectively, for A(Li), [C/Fe], [N/Fe], [O/Fe], [Na/Fe], [Mg/Fe], [Al/Fe], [Si/Fe], [Ca/Fe], [Ce/Fe], $[{\rm \langle Ba,La\rangle}/\rm Fe]$, and [(C+N+O)/Fe]. The associated error bars represent the standard deviation of the measurements. For comparison, we plot the second-order polynomial best fits to the lower and upper streams as solid and dashed gray lines, respectively.
The detailed abundance distributions for all considered elements, analogous to Figures~\ref{fig:chem_ls_us} and~\ref{fig:chem_agb}, are presented in Appendix~\ref{sec:ap1}.
Both the 2P and AII middle stream components occupy the region between the two gray lines for elements from Li to Al, while for [Si/Fe] both populations show lower-stream like values. No evident calcium differences are present between the three streams.
Within the limited [Fe/H] range covered by these stars, we do not detect significant differences between the median values of 2P and AII stars, suggesting no strong increasing or decreasing trends with [Fe/H]. A possible exception is [Na/Fe], for which the AII population shows a higher median value compared to 2P stars, although the two are consistent within 1$\sigma$.
The $R_{\rm s}$ values (in Appendix~\ref{sec:ap1}) confirm [Na/Fe] is the only light element showing a correlation with [Fe/H] ($R_{\rm s} = 0.54 \pm 0.12$). 

In the last three panels of Figure~\ref{fig:chem_middle}, we present the same analysis for the s-process elements [Ce/Fe], $[{\rm \langle Ba,La\rangle}/\rm Fe]$, and the total [(C+N+O)/Fe]. In all three cases, the average values for the middle stream tend to lie closer to those of the lower stream. However, the large dispersion prevents definitive conclusions. Nonetheless, the median values suggest potential increasing trends with [Fe/H], which are supported by their respective Spearman coefficients (see Appendix~\ref{sec:ap1}).

\section{The formation history of $\omega$Cen}
\label{sec:6}

In this section, we gather the observational evidence presented throughout this work in an effort to reconstruct the sequence of events that led to the fascinating complexity of $\omega$Cen. Sections~\ref{sec:6.1},~\ref{sec:6.2}, and~\ref{sec:6.3} explore the possible origins of the lower, upper, and middle streams, respectively, while Section~\ref{sec:6.4} summarizes the formation history that $\omega$Cen may have undergone.

\begin{figure}
\includegraphics[width=7cm, clip, trim={-3cm 0cm 0cm 0cm}]{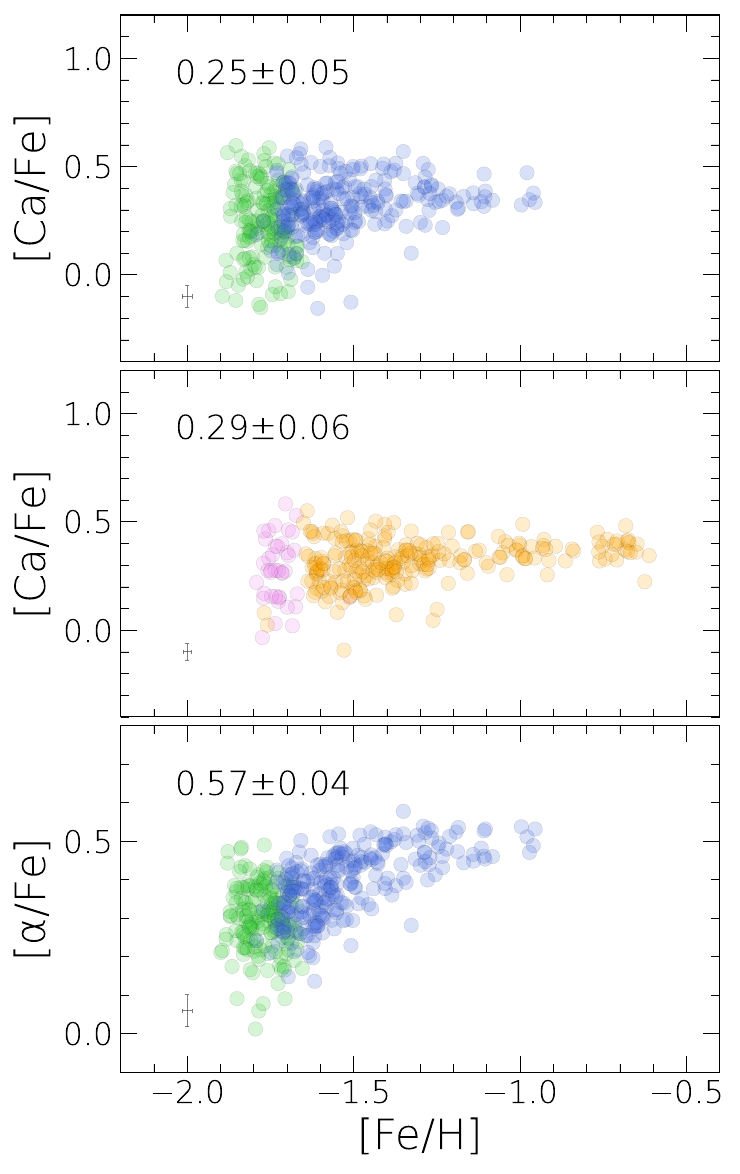}
\caption{Same as Figure~\ref{fig:chem_ls_us} but for the [Ca/Fe] of lower and upper stream stars (upper and middle panel, respectively), and [$\alpha$/Fe] vs. [Fe/H] of 1P and AI stars (lower panel).}
\label{fig:chem_alpha}
\end{figure}

\subsection{The lower stream}
\label{sec:6.1}

The 1P stars were likely the first generation to form in $\omega$Cen, originating directly from its primordial gas cloud. In nearly all proposed scenarios for multiple stellar populations, the 1P stars precede the 2P stars. The 1P lower metallicity compared to the anomalous populations \citep[see also Figure 19 from][]{johnson2010} strongly supports their earlier formation, as continued star formation typically leads to progressive iron enrichment.

A promising scenario for the formation of AI stars involves self-enrichment via core-collapse supernovae (CCSNe), originally proposed by \citet{marino2012} and recently supported by \citet{mason2025}. CCSNe produce $\alpha$-elements more efficiently than iron \citep[e.g.,][]{woosley1995, nomoto2006, marassi2019, boccioli2024}, potentially explaining the observed rise in [$\alpha$/Fe] with [Fe/H] in the lower stream (see Figure~\ref{fig:chem_alpha}). In this context, AI stars may have formed shortly after the 1P, from gas enriched by CCSNe ejecta.

In Figure~\ref{fig:yields}, we compare predicted CCSNe yields from a simple stellar population (SSP) to observed lower stream trends for [C/Fe], [N/Fe], [O/Fe], [Na/Fe], [Mg/Fe], [Al/Fe], [Si/Fe], [Ca/Fe], and [Ce/Fe] as functions of [Fe/H]. The yields were generated using the one-zone chemical evolution code {\tt{Chempy}} \citep{rybizki2017}, simulating CCSNe enrichment from a $10^{8}$~M$_{\odot}$ SSP with initial abundances matching 1P stars (black open dots), and assuming a \citet{kroupa2001} initial mass function. The simulation spans 40 Myr, corresponding to the typical timescale of CCSNe activity.
We emphasize that this is a qualitative, illustrative comparison; a detailed model-to-data fit is beyond this study’s scope. Our goal is to evaluate whether CCSNe yields align with the observed abundance trends.

The cumulative yields at 40 Myr are shown as a blue square and red triangle, representing two CCSNe yield sets: \citet[][LC2018]{limongi2018} and \citet[][N2013]{nomoto2013}, respectively. Both predict rising [Fe/H], [O/Fe], and [Si/Fe], consistent with observations. The N2013 yields capture the [Mg/Fe], [Al/Fe], and [Ca/Fe] trends, while LC2018 alone predicts a positive [C/Fe] trend—though much smaller than observed. The models also diverge significantly in [N/Fe]: LC2018 reproduces the observed flat trend, whereas N2013 shows a decline inconsistent with the data. Minor discrepancies are also seen in [Na/Fe].
These differences likely arise from the mass ranges adopted in the models: N2013 includes stars from 13–40 M${\odot}$, while LC2018 extends to 13–120 M${\odot}$. For [Ce/Fe]—available only from LC2018—the predicted increase is modest and fails to explain the observed substantial enrichment.

\begin{figure}
\includegraphics[width=9cm, clip, trim={0cm 0cm 25cm 0cm}]{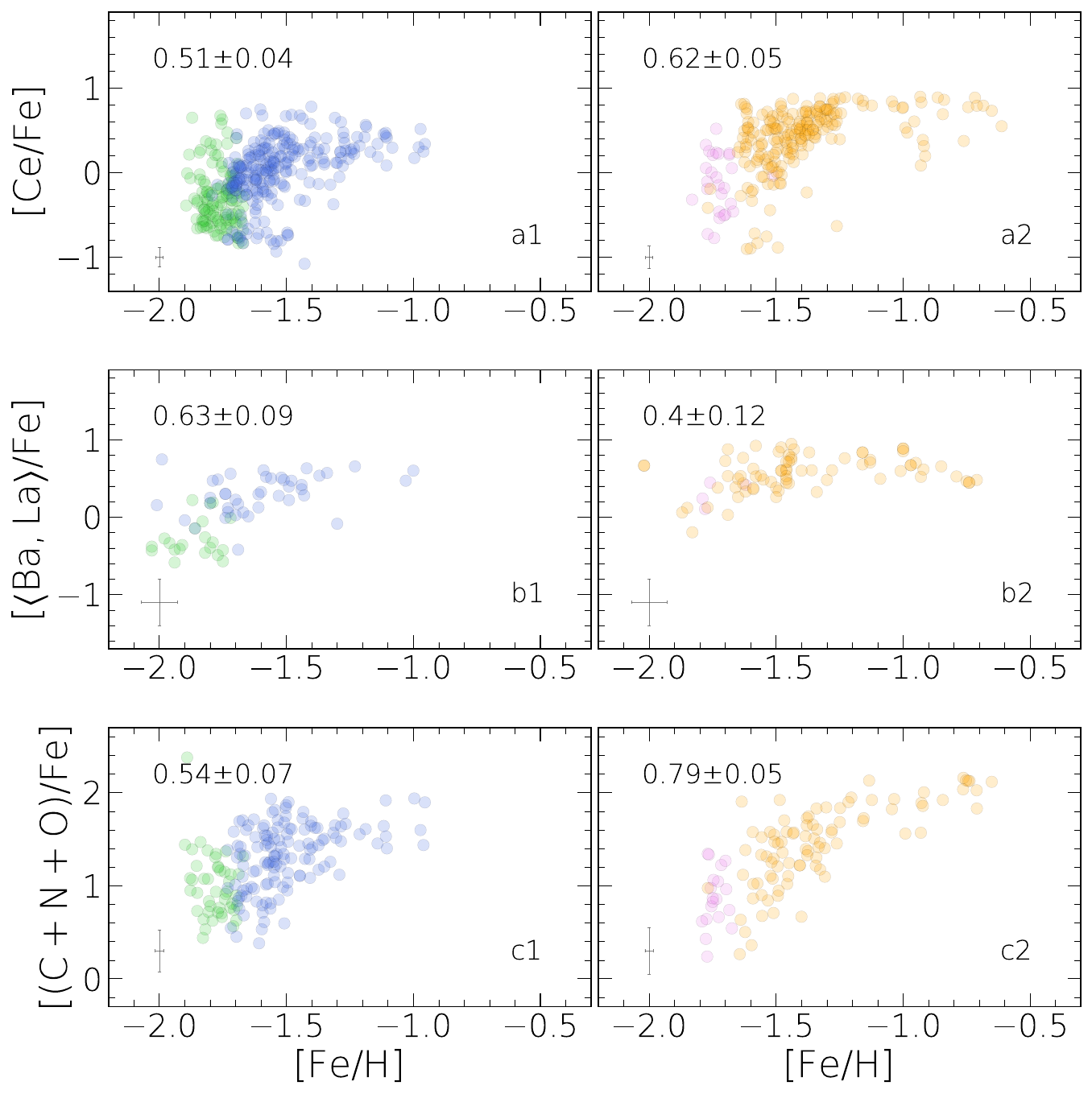}
\caption{Same as Figure~\ref{fig:chem_ls_us} but for the [Ce/Fe] (upper panels), $[{\rm \langle Ba,La\rangle}/\rm Fe]$ (middle panels), and [(C+N+O)/Fe] (lower panels) vs. [Fe/H] for lower and upper stream stars.}
\label{fig:chem_agb}
\end{figure}

\begin{figure*}
\includegraphics[width=18cm, clip, trim={0cm 0cm 0cm 0cm}]{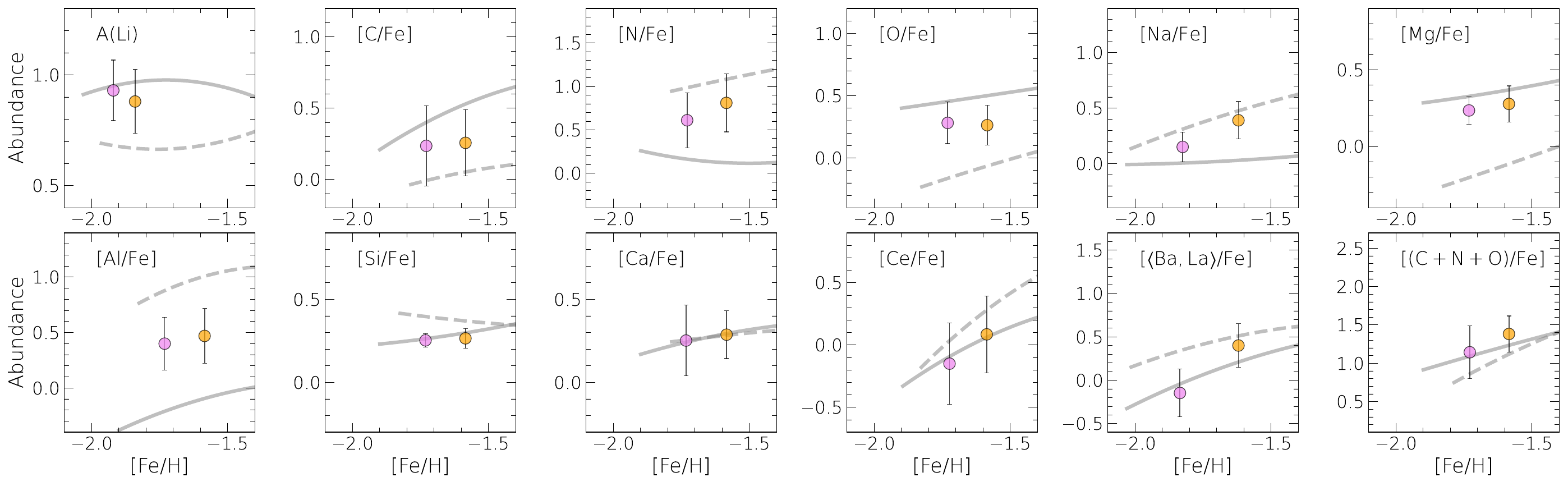}
\caption{Median A(Li), [C/Fe], [N/Fe], [O/Fe], [Na/Fe], [Mg/Fe], [Al/Fe], [Si/Fe], [Ca/Fe], [Ce/Fe], $[{\rm \langle Ba,La\rangle}/\rm Fe]$, and [(C+N+O)/Fe] vs. [Fe/H] of the 2P (violet) and AII (orange) middle stream stars. The continuous and dashed gray lines indicate the lower and upper stream average trend, respectively.}
\label{fig:chem_middle}
\end{figure*}

While the models do not match all abundance trends, they exhibit encouraging qualitative agreement, particularly for key $\alpha$-elements such as oxygen, magnesium, silicon, and calcium. These results support the idea that CCSNe played a major role in shaping the chemical evolution of the lower stream and make it a viable and promising formation scenario. 
Nonetheless, substantial mismatches remain—especially for carbon and cerium—highlighting the need for more detailed modeling and the contribution of other actors beyond CCSNe. An intriguing possibility involves stellar rotation: simulations by \citet{frischknecht2016} suggest that massive, rotating, metal-poor stars (15–40 M$_{\odot}$) can yield substantial s-process enrichment, thus explaining the large cerium increase.
Nevertheless, the chemical composition of massive stars' yields is a debated topic, with several factors, such as the  progenitor mass range, hypernovae contribution, and pre-supernovae yields, that heavily influence the chemical composition of the subsequent stellar generations \citep[see][for a recent review]{kobayashi2025}.

\subsection{The upper stream}
\label{sec:6.2}

The light-element chemical composition of the upper stream clearly indicates that these stars formed in a medium strongly polluted by the product of p-capture processes, exhibiting the signature lithium, carbon, oxygen, and magnesium depletion and nitrogen, sodium, and aluminum enhancement.
In the multiple population formation scenarios, 2P stars with the most extreme chemical composition are believed to be the first one formed after the 1P from pure polluters' ejecta.
\citet{marino2012} proposed to extend this idea to $\omega$Cen, with lower stream stars at different [Fe/H] ejecting the material from which the upper stream originated, as a natural extension of the multiple population theory.
Our observation of AII having similar radial segregation than 2P stars supports this idea, as it is a common feature of the multiple populations phenomenon.

In a simple model where the upper stream formed from the chemically processed ejecta of lower stream stars, we would expect similar abundance trends with [Fe/H], simply shifted depending on the element’s production or destruction during the p-capture processes. However, Figure~\ref{fig:chem_ls_us} shows a far more complex behavior. Moreover, the upper stream reaches [Fe/H] up to $\sim$-0.6 dex, but no lower stream stars exhibit such high value. These facts seems difficult to conciliate to the idea that the upper stream formed from lower stream ejecta, but can be justified by considering additional physical processes.

First, the production of magnesium, aluminum, and silicon during proton-capture reactions is highly metallicity-dependent. At lower [Fe/H], higher core temperatures in polluter stars ($>$70 MK) ignite the Mg–Al cycle, depleting Mg and enhancing Al. At even higher temperatures ($>$100 MK), silicon is produced via leakage from this cycle \citep[][]{arnould1999, prantzos2017}. Thus, more metal-poor stars —being hotter— undergo stronger Mg depletion and Al/Si enrichment, a pattern confirmed in several observations \citep[e.g.,][]{pancino2017, dondoglio2025}.
This explains the steeper [Mg/Fe] decline with [Fe/H] seen in the upper stream: Mg depletion lessens as metallicity rises. The arc-shaped trend in [Al/Fe] follows a similar logic. Between [Fe/H] $\sim$ –1.7 and –1.4 dex, [Al/Fe] increases, echoing the behavior of its lower stream progenitors. At higher [Fe/H], Al enhancement drops sharply, reversing the slope in the [Al/Fe]–[Fe/H] plane. The decline in [Si/Fe] in the upper stream shares the same physical origin: since silicon production requires even higher temperatures than aluminum, its downturn begins at lower metallicities and thus lacks the initial upturn observed in Al. This metallicity dependence constitutes a strong proof that the p-capture processed material that contributed in building Fe-rich AII stars was ejected by stars that are metal-richer than the 1P stars, with AI being the most logical candidates.

Second, additional sources of chemical enrichment may contribute. As increasingly metal-rich lower stream stars pollute the medium, earlier generations (such as 1P stars) evolve and release ejecta from intermediate-mass AGB stars (3–4 M$_\odot$) and Type Ia supernovae. \citet{dantona2016} proposed these sources to explain chemical anomalies in Type II GCs. Their material could mix with p-capture products and contribute to upper stream formation, explaining the rise in [N/Fe] and [Na/Fe], both expected from AGB yields \citep{ventura2013, karakas2014}, and the higher [Fe/H] values in the upper stream, as supernovae Ia are prolific iron producers \citep[e.g.,][]{gronow2021, keegans2023}.
In particular, 3–4 M$_\odot$ AGB stars are thought to be major contributors of s-process elements \citep[e.g.,][]{karakas2014, cristallo2015}. If such stars influenced the upper stream’s formation, we would expect higher s-process abundances compared to the lower stream. This is supported by the [Ce/Fe] distributions in Figure~\ref{fig:chem_agb}, which show higher values in the high-[Fe/H] end of the upper stream.

Our observations are in agreement with the idea that, similar to what hypothesized for Type I GCs but extended to a structure with a longer star formation history, ejecta from lower stream significantly contribute in forming upper stream stars.

\begin{figure*}
\includegraphics[width=18cm, clip, trim={0cm 0cm 0cm 0cm}]{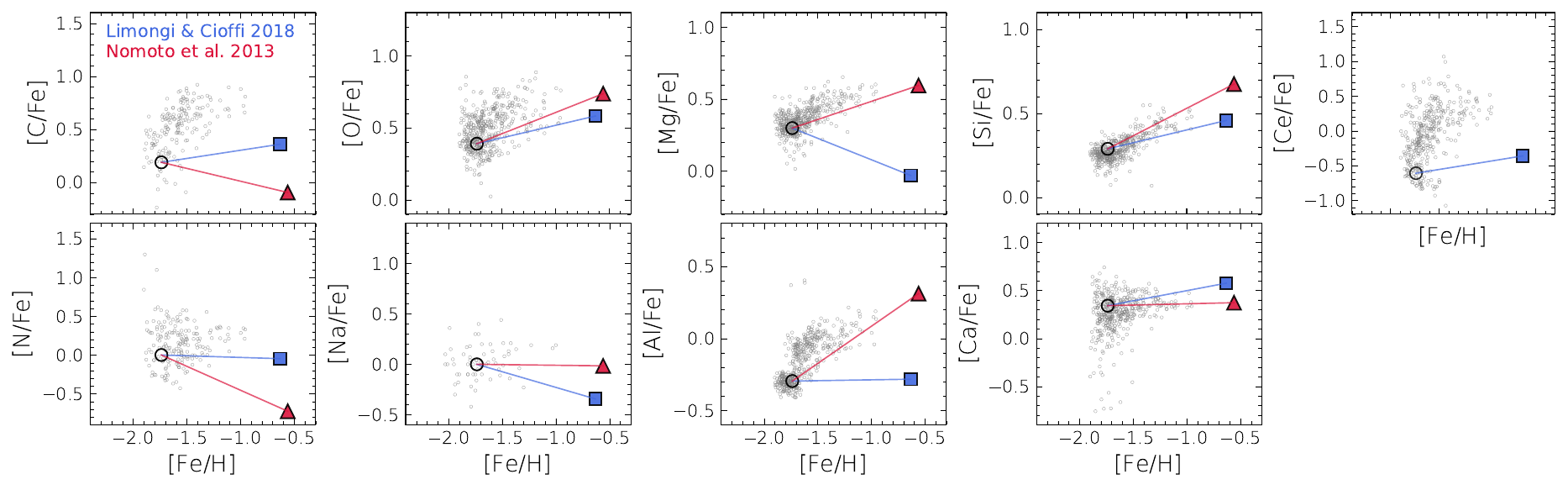}
\caption{Distribution of nine abundance ratios (see labels) vs. [Fe/H] of the lower stream stars compared with yields produced by CCSNe from a simulated population with initial composition indicated by the open black dots. The blue square and the red triangle indicate yields produced using the \cite{limongi2018} and \citet{nomoto2013} tables.
}
\label{fig:yields}
\end{figure*}

\subsection{The middle stream}
\label{sec:6.3}

The middle stream offers arguably the most enigmatic chemical pattern of the three, with its much smaller [Fe/H] spread constituting one of the most challenging piece to fit in solving the $\omega$Cen puzzle.

The 2P stars in the middle stream offer a useful starting point for interpreting the origin of this stellar component. An extended 2P with a wide range of light-element abundances is a common feature of massive Galactic GCs. A well-studied example is NGC\,2808, which hosts a fraction of chemically extreme 2P stars, along with a more extended group of 2P stars whose compositions gradually resemble 1P stars \citep[][]{carretta2015, milone2015, carlos2023}.

A popular explanation for this phenomenon involves a dilution mechanism \citep[e.g.,][]{dantona2016, carretta2018}. In this scenario, the most extreme 2P stars form after the 1P from nearly pure ejecta in the dense central regions of a GC. Later, the intracluster medium —with 1P-like chemical composition— flows into the core and mixes with the residual ejecta, producing new 2P stars. The more this dilution occurs, the more the resulting 2P stars resemble 1P stars in their chemical signatures.
The intermediate chemical composition of middle stream 2P stars supports the dilution scenario, suggesting that these stars formed from a mixture of polluters’ ejecta and 1P-like material.

\begin{figure*}
\includegraphics[width=17cm, clip, trim={0cm 0cm 0cm 0cm}]{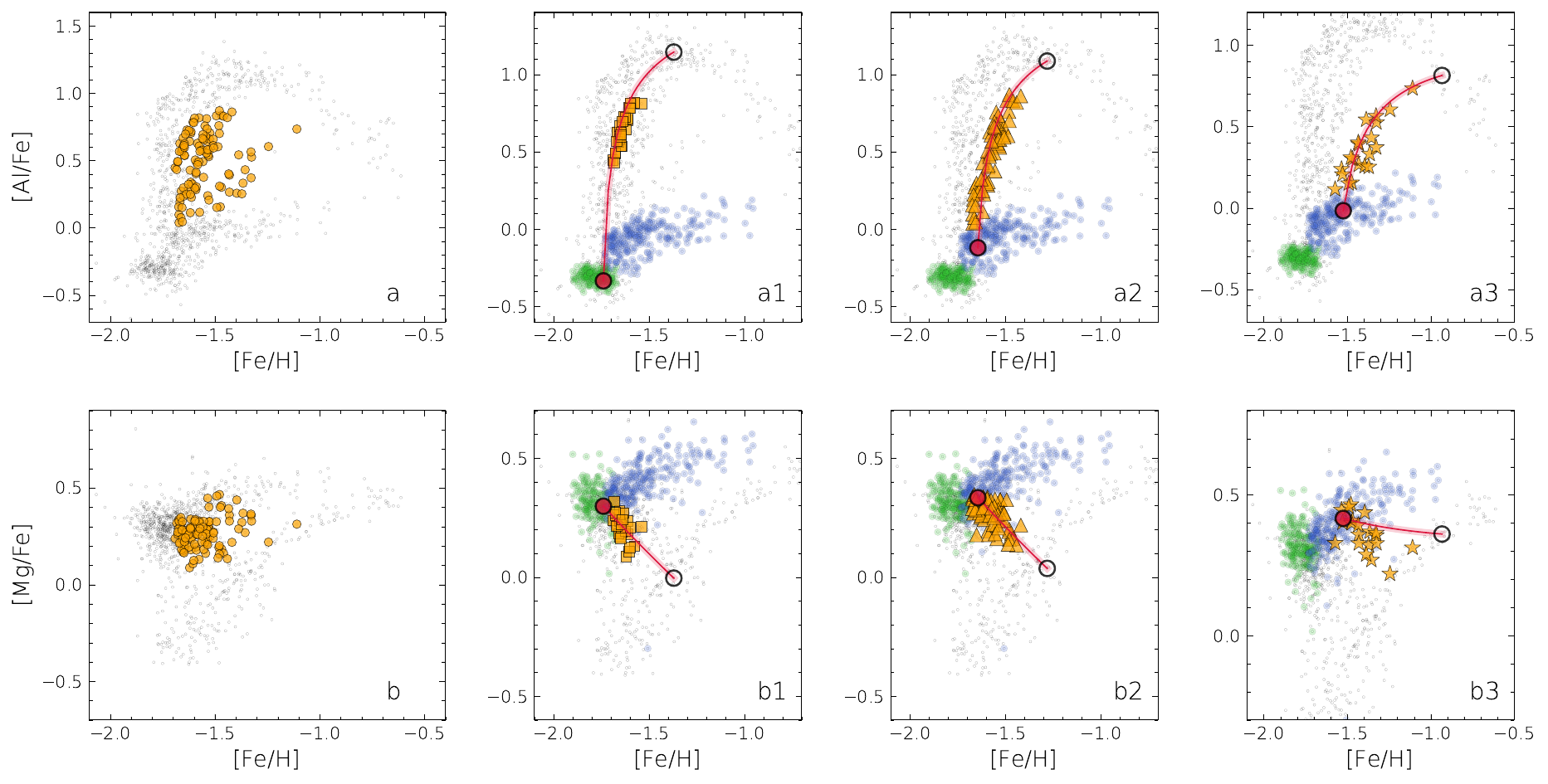}
\caption{[Al/Fe] vs. [Fe/H] diagram with the middle stream AII stars highlighted with orange bullets (Panel a). Panels a1, a2, and a3 represent the same plot zoomed-in around seq1, seq2, and seq3, respectively (see text), with lower stream stars highlighted for reference. The best-fit fiducial lines are represented in red, with the empty and filled red points representing the 0\% and 100\% dilution points, respectively.
Panels b, b1, b2, and b3, show the same but in the [Mg/Fe] vs, [Fe/H] plane.}
\label{fig:aII_sequences}
\end{figure*}

But can dilution also explain the anomalous middle stream?
In panel a of Figure~\ref{fig:aII_sequences}, we highlight the AII stars associated with the middle stream (in orange) on the [Al/Fe] vs. [Fe/H] diagram. They define three distinct sequences: Sequence 1 (seq1), from ([Fe/H], [Al/Fe])$\sim$(–1.65, 0.55) to (–1.55, 0.80), marked with squares in panel a1, Sequence 2 (seq2), from (–1.65, 0.05) to (–1.45, 0.80), shown with triangles in panel a2, and Sequence 3 (seq3), from (–1.55, 0.10) to (–1.10, 0.80), indicated by starred symbols in panel a3.
In all three sequences, [Al/Fe] drops as [Fe/H] decreases, a trend suggestive of dilution between material enriched in Al (from upper stream stars) and a gas that is both Al- and Fe-poorer. Analogous to Type I GCs, where 2P material is diluted by 1P-like gas, the most natural diluting agent here is gas resembling that of the lower stream.

To test this, we computed theoretical dilution curves as follows: (i) we selected upper stream stars as the undiluted end-members and converted their [Fe/H] and [Al/Fe] to mass fractions. (ii) We mixed these with increasing fraction of lower-stream-like material and calculated the resulting Fe and Al mass fractions. (iii) The diluted values are then converted back to abundance ratios for comparison in the [Al/Fe] vs. [Fe/H] plane.
The dilution curves that qualitatively fit seq1, seq2, and seq3 are overplotted in red in panels a1, a2, and a3, respectively. Filled-red and empty circles indicate the 100\% and 0\% dilution points. Lower stream stars are highlighted as in Section~\ref{sec:3} for reference.
In panels b, b1, b2, and b3, we repeated the analysis in the [Mg/Fe] vs. [Fe/H] plane to assess whether the derived dilution curves also align with the magnesium trends of the middle-stream anomalous stars.

We find that seq1 is consistent with dilution of upper stream material with [Fe/H]$\sim$–1.40 by 1P-like gas (panel a1). The corresponding curve in the [Mg/Fe] vs. [Fe/H] diagram (panel b1) also aligns with the observed data. Stars in seq1 span only part of the predicted dilution track, corresponding, in both [Al/Fe] and [Mg/Fe] distributions, to a $\sim$65–90\% range of dilution.
For seq2, the starting point is at higher [Fe/H] ($\sim$–1.25 dex), and the dilution agent must be more metal-rich than typical 1P gas ([Fe/H]$\sim$–1.60). Dilution with 1P-like gas fails to reproduce the observed trends, especially in [Mg/Fe] vs. [Fe/H] (panel b2), where the sequence clearly moves toward the AI region of the diagram. As in seq1, data in both plots span a similar dilution range ($\sim$55–100\%).
Finally, seq3 shows inconsistency between its Al and Mg trends. While both panels a3 and b3 indicate dilution with lower-stream gas of [Fe/H]$\sim$–1.55, the initial (0\% dilution) point that fits the Al trend ([Fe/H]$\sim$–0.95) does not match the Mg trend, which requires a Fe-richer end by $\sim$0.20 dex.

Our qualitative analysis indicates that dilution is a viable explanation for the middle stream, at least for two of the three identified sequences. The corresponding requirement for an increasingly iron-rich diluting agent to reproduce sequences with higher [Fe/H] supports our hypothesis (Section~\ref{sec:6.3}) that the iron-rich upper stream stars formed in an environment progressively enriched in iron—likely by stars along the AI sequence.

\subsection{Putting the pieces together}
\label{sec:6.4}

Based on the observational results presented in Sections~\ref{sec:3}–\ref{sec:5} and the discussion in Sections~\ref{sec:6.1}–\ref{sec:6.3}, we propose a possible star formation sequence that led to $\omega$Cen.

The 1P stars formed first from the primordial molecular cloud. Unlike typical GCs, $\omega$Cen’s deep potential well retained CCSNe ejecta, enabling the formation of AI stars from increasingly enriched gas—likely facilitated by its early role as part of a more massive system, such as a dwarf galaxy.
Later, 1P massive stars (depending on the nature of the multiple populations polluter) polluted the central regions with material enriched in p-capture elements, forming from pure ejecta upper stream 2P stars. As 1P-like gas flowed inward, it diluted this ejecta, producing 2P stars with compositions closer to the 1P population \citep[see][]{dantona2016}. The radial distribution of these stars (Figure~\ref{fig:radial}) preserves this evolutionary imprint.
A similar process occurred for AI stars, which evolved later. Their ejecta formed the upper stream AII population, progressively enriched in Fe and $\alpha$-elements, but also affected by pollution from 1P AGB stars ($\sim$3–4 M$_{\odot}$) and Type Ia SNe. These events increased Fe and s-process elements in upper stream stars (Figure~\ref{fig:chem_agb}). As with 2P stars, dilution with lower-stream-like gas occurred, reducing not only light-elements differences but also decreased the [Fe/H], since the diluting gas was Fe-poor compared to the newly forming stars.
Table~\ref{tab:origin} summarizes the formation timeline, chemical features, and proposed origins of these populations.

The timescale at which this whole formation history might occur (i.e., the age spread of $\omega$Cen) depends on several assumptions regarding the different polluters involved, such as the mass range of exploding stars in the CCSNe phase, the Type Ia supernovae rate in a $\omega$Cen like environment, and the AGB nucleosynthesis assumptions in different mass ranges. An in-depth modeling of these phenomena is beyond the scope of the current work, which focuses on providing observational constrains.
However, we point out, based on the scheme proposed by \citet[][see their table 4]{dantona2016}, that the CCSNe phase should end $\sim$40 Myr after the formation of the 1P, while the intermediate-mass AGB and Type Ia supernovae are expected to start polluting the medium after $\sim$100 My.
Moreover, after enough time \citep[$\sim$300-400 Myr,][]{ventura2013} also the ejecta from lower-mass AGB stars ($<$3M$_{\rm \odot}$) start pollute the cluster medium. These stars mostly produce a carbon enrichment, which is only observed in the lower stream. However, pollution from these sources cannot increase heavier elements such as magnesium, aluminum and sodium \citep[][and references therein]{ventura2022}, thus with CCSNe ejecta being a much more realistic candidate for the formation of these stars.
Consequently, the star formation history in $\omega$Cen would have ended before $<$3M$_{\rm \odot}$ AGB stars strongly polluted the medium.
These time scales suggest that the formation of $\omega$Cen lasted a few hundreds of Myr, in agreement with the estimate from \citet{tailo2016}, who posed 500 Myr as the maximum age spread, but at odds with other works in the literature, which derived a $\sim$2 Gyr age spread \cite[e.g.,][]{villanova2014, clontz2024}.

\section{Summary and Conclusions}
\label{sec:7}

In our work, we combined multiple datasets to consistently track different $\omega$Cen populations from the cluster core out to $\sim$30 arcmin, focusing on RGB stars. Our main findings are summarized below:

\begin{itemize}
\item The combination of {\it{HST}} multi-band photometry and spectroscopic abundances enabled us to identify the canonical 1P and 2P, as well as the anomalous AI and AII populations, over an unprecedented radial extent—approximately five times the half-light radius. This approach ensures consistency between photometric and spectroscopic tagging.

\item This population tagging allowed us to investigate their radial distributions. We found that 2P stars are more centrally concentrated than 1P stars, as detected in other GCs. Moreover, the spatial distributions of the AI and AII populations mirror those of the 1P and 2P stars, respectively, with AII stars being more spatially concentrated than AI stars. Interestingly, binary stars are more prevalent among the 1P and AI populations, indicating a higher binary fraction compared to their chemically extreme counterparts.

\item Our dataset allowed a deep investigation of the chemical variety within the three streams and a their overall differences spanning the whole $\omega$Cen's [Fe/H] range. In particular, we provide robust constrain on the C+N+O content of each stream and their relation with metallicity, based -for the first time- only on stars below the RGB bump, thus not affected by mixing episodes.

\item A comparison between the lower stream (1P and AI stars) and the upper stream (chemically extreme 2P and AII stars) supports a scenario in which the upper stream formed from gas polluted by products of p-capture processes, synthesized and expelled into the intracluster medium by lower stream stars. However, to fully explain all the observed chemical features, additional polluting mechanism must be invoked.

\item Self enrichment from CCSNe ejecta constitutes the most promising mechanism behind the chemical inhomogeneities among lower stream stars. However, SSP yield models fail to reproduce the observed increases in certain abundance ratios -particularly for carbon and cerium- as [Fe/H] increases. Additional mechanisms, such stellar rotation in massive stars, may justify (at least) some of the discrepancies.

\item Starting from multiple-population scenarios that include dilution, we propose a similar mechanism to explain the chemical morphology of AII stars in the middle stream. In the [Al/Fe] vs. [Fe/H] diagram, three distinct sequences are evident, which we interpret as dilution curves between Al- and Fe-rich gas and the lower stream-like material, converging toward smaller [Fe/H] and [Al/Fe] values.

\end{itemize}

Based on the observational evidence gathered in our study, we propose a plausible sequence of events leading to the complex structure of $\omega$Cen. While the precise chronology depends on the origin of the proton-capture processed material and the timescales of the different polluters involved, the underlying scenario is the following: the lower stream stars pollute the intracluster medium with their ejecta, from which the upper stream stars subsequently form. Later, the middle stream emerges through the dilution of residual polluted gas (not collapsed into upper stream stars) with the intracluster material with chemical composition typical of lower stream stars.
Notably, our scheme, based on spatial distribution and chemical composition only, is in overall agreement with kinematics investigations: upper stream stars exhibit significant radial anisotropy, while lower stream stars are nearly isotropic, a difference typically observed between the 1P and 2P of dynamically young GCs \citep{cordoni2020, ziliotto2025}, thus providing further, independent support that the whole upper and lower stream arose from a mechanism similar to the one producing 1P and 2P stars.

In this scenario, $\omega$Cen would form by self-enrichment processes, without the necessity of invoking merging events from several GCs. In this case, this cluster would constitute an extension of the so-called Type II GCs (where variation in iron, s-process elements and total C+N+O are observed) that originated from a more massive structure, likely a dwarf galaxy, allowing to retain a larger amount of massive stars ejecta and thus exhibiting a prolonged star formation history.

\begin{acknowledgements}
This work has been funded by the European Union – NextGenerationEU RRF M4C2 1.1 (PRIN 2022 2022MMEB9W: “Understanding the formation of globular clusters with their multiple stellar generations”, CUP C53D23001200006).
T. Ziliotto acknowledges funding from the European Union’s Horizon 2020 research and innovation programme under the Marie Skłodowska-Curie Grant Agreement No. 101034319 and from the European Union – NextGenerationEU".

\end{acknowledgements}

\bibliographystyle{aa}
\bibliography{aanda}

\begin{appendix}

\section{Details on our spectrosocpic dataset} \label{sec:ap0}

We dedicate this appendix to provide more details about the spectrosocpic datasets employied in this work, as follow:

\begin{itemize}

    \item APOGEE. This survey, carried out using the du Pont Telescope and the Sloan Foundation 2.5 m Telescope \citep{gunn2006} at Apache Point Observatory, observed over 700,000 stars at a spectral resolution of R$\sim$22,500. DR17 provides detailed chemical abundance information for multiple stellar populations across 21 GCs in our sample, representing the most comprehensive spectroscopic dataset overlapping with our photometric sample.
    From this release, we selected APOGEE stars with signal-to-noise ratios greater than 70 and excluded those with the {\tt{ASPCAPFLAG}} marked as {\tt{STAR\_BAD}}, which indicates issues such as numerous bad pixels or a non-stellar classification. As noted by \citet{jonsson2020}, APOGEE measurements of [Na/Fe] are based on relatively weak spectral lines, making sodium one of the least reliably measured elements in this survey. Given the extensive sodium abundance data available in the literature and the limitations of APOGEE for this element, we exclude [Na/Fe] values derived from APOGEE spectra from our analysis.

    \item \citet{marino2011}.
    The dataset comprises a substantial collection of FLAMES/GIRAFFE at the Very Large Telescope spectra for approximately 300 giant stars, each with a signal-to-noise ratio between 60 and 100. Observations were obtained using four instrumental setups (HR09B, HR11, HR13, and HR15) with a spectral resolution ranging from R$\sim$20,000 to 25,000. We utilized these spectra to extract measurements of [Na/Fe], [Ba/Fe], and [La/Fe]. Sodium abundances were derived from the doublet near 6150 \AA, barium from the blended BaII line at 6141 \AA, and lanthanum from the LaII lines at 6262 \AA and 6390\AA.
    
    \item \citet{mucciarelli2018}. Observations were carried out using both the FLAMES multi-object spectrograph and UVES, targeting approximately 200 giant stars in $\omega$Cen located below the red giant branch (RGB) bump. The signal-to-noise ratio and spectral resolution are comparable to those reported by \citet{marino2011}. From this dataset, we utilized A(Li) measurements derived from the lithium line at 6708 \AA.
    
\end{itemize}

\section{Middle stream abundances vs [Fe/H]} \label{sec:ap1}

In this appendix, we show the analogous of Figures~\ref{fig:chem_ls_us} and~\ref{fig:chem_agb} but for the middle stream stars. As done in Section~\ref{sec:5}, colored in violet and orange according to their belonging to the 2P and AII populations, respectively.
[Na/Fe], [Ce/Fe], $[{\rm \langle Ba,La\rangle}/\rm Fe]$, and [(C+N+O)/Fe] show $R_{\rm s}$ consistent with a correlation with [Fe/H], while no other abundance displays any evident trend with the iron over hydrogen ratio. \\\\

\begin{figure*}
\includegraphics[width=17cm, clip, trim={0cm 0cm 0cm 0cm}]{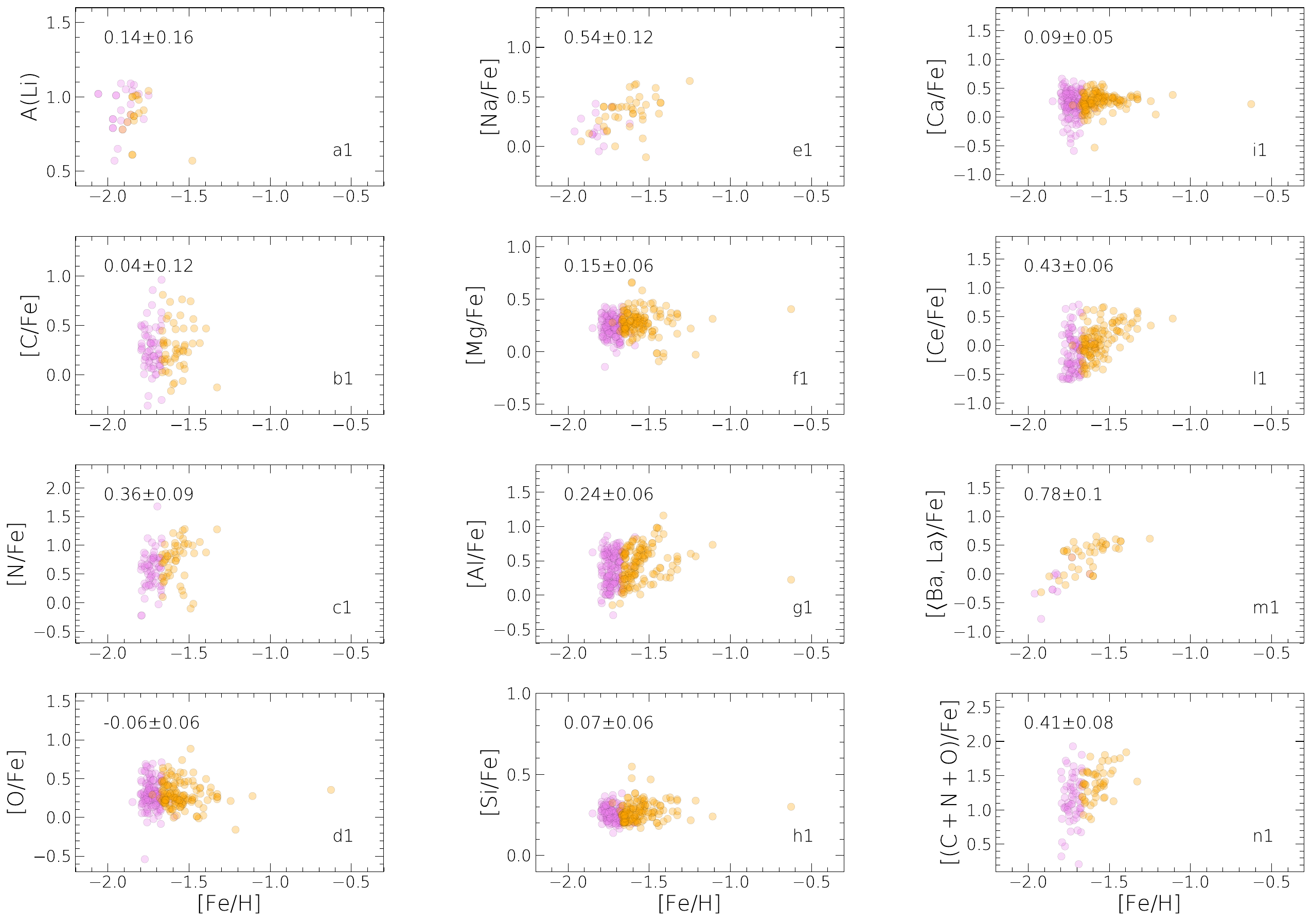}
\caption{A(Li), [C/Fe], [N/Fe], [O/Fe], [Na/Fe], [Mg/Fe], [Al/Fe], [Ca/Fe], [Si/Fe], [Ce/Fe], $[{\rm \langle Ba,La\rangle}/\rm Fe]$, and [(C+N+O)/Fe] vs. [Fe/H] of the middle stream. Violet and orange dots indicate AI and AII stars, respectively. The numbers reported in the panels illustrate the Spearman's correlation coefficient and its associated error. }
\label{fig:app_ms}
\end{figure*}

\section{Table scheme of the chemical evolution of $\omega$Cen} \label{sec:ap2}

In this Appendix, we provide a table that summarizes the proposed scenario in Section~\ref{sec:6}. The identified populations are ordered chronologically from the first to the last row. The Table also provides a summary of the observed chemical features and their origin according to our scenario.

\begin{table*}
\caption{Summary of the timeline, the chemical features, and the origin of the different populations in $\omega$Cen.}
\centering
\begin{tabular}{p{0.1cm}p{2.8cm}|p{5.8cm}|p{5.8cm}}
\hline
\hline
& & & \\
& Timeline   & Chemical features & Origin \\
& & & \\
\hline
 & & & \\
 & & & \\
\multirow{6}{*}{%
  \begin{tikzpicture}[baseline=(current bounding box.center)]
    \draw[->, line width=0.5pt, >=latex] (2,0) -- (2,-12cm);
  \end{tikzpicture}}

& 1P                  & Similar to Galactic halo stars, with a small iron spread.                                          & Formed from a primordial cloud. \\
& & & \\
& AI                  &
Enriched in C+N+O (primarily due to C and O), $\alpha$-elements, iron, and s-process elements, relative to 1P stars.      & Medium polluted by ejecta from CCSNe (and possibly other massive stars). \\
& & & \\
& & & \\
& 2P (upper stream)   & Depleted in Li, C, O, and Mg; enriched in N, Na, Al, and Si relative to 1P stars. No major differences in heavier elements between 1P and 2P.
& Medium composed of pure ejecta from p-capture processes associated with multiple-population polluters. \\
& & & \\
& & & \\
& 2P (middle stream)  & Intermediate levels of C, N, O, Na, and Al between upper-stream 2P and 1P stars. Li, Mg, and Si are consistent with 1P abundances.
& Result of dilution between pure 2P ejecta and a primordial, 1P-like medium. \\
& & & \\
& & & \\
& AII (upper stream)  & Light-element patterns similar to those in AI stars compared to 1P–2P differences. Covers a broad [Fe/H] range. N, O, Na, Mg, Al, s-process elements, and C+N+O increase with [Fe/H]; Si decreases.
& Formed from a mixture of polluters: p-capture ejecta from AI stars, 3–4M$_{\rm \odot}$ AGB stars, and SN Ia. Higher [Fe/H] stars result from material ejected by iron-richer sources. \\
& & & \\
& & & \\
& AII (middle stream) &
Light-element abundances are intermediate between upper-stream AII and AI stars, similar to middle-stream 2P stars. Shows a smaller iron spread than anomalous stars in the upper and lower streams. [Fe/H] decreases toward the lower stream.
& Originates from dilution between upper-stream AII-like gas and lower-stream-like material with lower average [Fe/H].
\\
& & & \\
& & & \\
\hline
\hline
\end{tabular}
\label{tab:origin}
\end{table*}

\end{appendix}

\end{document}